\documentstyle[epsfig]{mn2e}

\def\gtsima{$\; \buildrel > \over \sim \;$}
\def\ltsima{$\; \buildrel < \over \sim \;$}
\def\gtrsim{\lower.5ex\hbox{\gtsima}}
\def\lesssim{\lower.5ex\hbox{\ltsima}}

\newcommand{\ergs}{\rm erg\,s^{-1}}

\begin{document}

\title[X-ray ionized nova ejecta in RZ 2109?]{Broad [OIII] in the globular cluster RZ~2109: X-ray ionized nova ejecta?}
\author[Ripamonti et al.]
{E. Ripamonti$^{1}$, M. Mapelli$^{2}$
\\
$^1$Universit\`a di Milano-Bicocca, Dipartimento di Fisica G.Occhialini, Piazza
delle Scienze 3, I--20126, Milano, Italy\\
$^2$INAF - Osservatorio Astronomico di Padova, Vicolo dell'Osservatorio 5, I-35122, Padova, Italy
}
\maketitle \vspace {7cm }

\begin{abstract}
We study the possibility that the very broad ($\sim$ 1500 km s$^{-1}$)
and luminous ($L_{5007} \sim 1.4 \times10^{37}\,{\rm erg\,s^{-1}}$)
[OIII] line emission observed in the globular cluster RZ~2109 might be
explained with the photoionization of nova ejecta by the bright
($L_{\rm X}\sim 4\times10^{39}\,{\rm erg\,s^{-1}}$) X-ray source
hosted in the same globular cluster. We find that such scenario is
plausible and explains most of the features of the RZ~2109 spectrum
(line luminosity, absence of H emission lines, peculiar asymmetry of
the line profile); on the other hand, it requires the nova ejecta to
be relatively massive ($\gtrsim 0.5\times10^{-3}\,{\rm M_\odot}$), and
the nova to be located at a distance $\lesssim 0.1\, {\rm pc}$ from
the X-ray source. We also predict the time evolution of the RZ~2109
line emission, so that future observations can be used to test this
scenario.
\end{abstract}

%
%________________________________________________________________

\section{Introduction}
Many X-ray sources (XRSs) are inside globular clusters (GCs;
e.g. Heinke 2011), but only five of them unambiguosly host a black
hole (BH; Maccarone et al. 2007, hereafter M07; Brassington et
al. 2010; Shih et al. 2010; Maccarone et al. 2011; Irwin et al. 2010,
hereafter I10; see King 2011 for an alternative explanation). These
five objects were observed as luminous XRSs ($L_{\rm
  X}\gtrsim 4 \times 10^{38}\, \ergs$)
%(point-like non-nuclear objects with $L_{\rm X}\ge10^{39}\,\ergs$)
with strong
variability (so that blending is excluded; Kalogera et al. 2004).
% positionally
%coincident with GCs in the elliptical galaxies NGC 4472, NGC 1399, and
%NGC 3379 (check), whose

The spectra of at least two of their host GCs are quite peculiar. One
of them, RZ~2109 (in the elliptical NGC~4472; Zepf et al. 2007, 2008,
and Steele et al. 2011; hereafter Z07, Z08, S11) displays prominent
and broad [OIII]$\lambda\lambda$5007,4959 emission (the FWHM of
[OIII]$\lambda5007$ is $1000$--$2000\,{\rm km\,s^{-1}}$; its flux is
$F_{5007} \sim 4.4 \times 10^{-16}\,{\rm erg\, s^{-1}\, cm^{-2}}$; at
$D = 16\, {\rm Mpc}$ it corresponds to $L_{5007} \sim 1.4
\times10^{37}\,{\rm erg\,s^{-1}}$; see Z08), but no H line
emission\footnote{No other emission line (e.g.  [OII]$\lambda3727$,
  [NII]$\lambda\lambda6548,6583$, [SII]$\lambda\lambda6716,6731$) is
  detected, except perhaps [OIII]$\lambda4363$.
%We also note that Z07 estimated a much lower $F_{5007}\simeq0.13
%\times 10^{-16}\,{\rm erg\, s^{-1}\, cm^{-2}}$ (but they warn that
%they might underestimate this flux by a factor of $\sim 2$), but this
%difference might be explained by the different observational setups.
}.  The spectrum of the GC hosting the XRS CXOJ033831.8-352604 (in the
elliptical NGC~1399; see I10) is somewhat similar in the absence of H
and the presence of [OIII] lines; but widths and luminosities are
lower (FWHM $\simeq 140\,{\rm km\,s^{-1}}$, $L_{5007}\sim {\rm
few}\times10^{36}\,\ergs$), and [NII] $\lambda\lambda$6583,6548
emission is comparable to [OIII].

These observations are hard to explain, because of the high
[OIII]/H$\beta$ ratio (Z08 estimate $\sim 30$; I10 finds a 3$\sigma$
lower limit of 5; both are very uncertain) and of the large
broadening, implying motions much faster than GC velocity dispersions
($\sim 10\,{\rm km\,s^{-1}}$). Optical lines are too wide for a
planetary nebula, and too narrow for a supernova remnant. Z08 suggest
that the [OIII] emission from RZ~2109 might come from a strong wind
originating around the BH, photoionized by the XRS. Such scenario
strongly favours a BH with mass $\approx10{\rm M_\odot}$, rather than
an intermediate-mass BH (IMBH) with mass in the 10$^2$-10$^5$M$_\odot$
range. This was preferred to the hypothesis that the [OIII]-emitting
gas closely orbits ($r\sim10^{13}\,{\rm cm}$) an IMBH, on the basis of
considerations about the maximum possible emission from such a small
volume. Such argument is severely weakened if the orbiting gas is very
metal-rich, e.g. as the result of the tidal disruption of a white
dwarf (WD) by an IMBH (see I10): according to Porter (2010), the
CXOJ033831.8-352604 data are compatible with this scenario,
% (though the [NII] lines might be problematic)
whereas in the case of RZ~2109 the conflict remains; however, Clausen
\& Eracleous (2011) showed that simulations of WD disruption
can reproduce the [OIII] line profiles observed in RZ~2109. On the
other hand, Maccarone \& Warner (2011) argued that the line emission
of CXOJ033831.8-352604 might be better explained with the
photoionization (by the XRS) of the wind from an R Coronae Borealis
star in the same GC.

In this paper, we examine a further possible scenario, and apply it to
the case of RZ~2109: [OIII] lines might be associated with a recent
nova eruption in the core of the GC. In particular, we discuss the
case (first mentioned by S11) where the [OIII] emission comes from the
photoionization of the nova ejecta by the XRS. X-ray emission from
novae is typically fainter ($L_{\rm X} \sim 10^{38}\,{\rm
  erg\,s^{-1}}$, even though at least one nova was reported to reach
$L_{\rm X} \gtrsim 10^{39}\,{\rm erg\,s^{-1}}$ - see Henze et
al. 2010) and shorter (decay times $\lesssim 1 {\rm yr}$) than what is
observed for these sources. Thus, we assume that the nova eruption is
{\it not} associated to the XRS, i.e. that the co-existence of the two
phenomena in the same GC is serendipitous.
%; however, we also discuss
%the possibility that the nova and the XRS might be physically related.

\section{Novae}
Nova eruptions occur in binary systems where a WD accretes from a
normal star (e.g. Gehrz et al. 1998). Nuclear reactions ignite
explosively in the material accumulated on the WD surface, leading to
a brief super-Eddington phase during which the WD outer layers are
ejected (e.g. Starrfield et al. 1972, Prialnik 1986). Ejecta
velocities are in the 100--2500$\,{\rm km\,s^{-1}}$ range, and their
metal mass fractions can be as high as $Z\sim 0.5$ (Yaron et al. 2005,
Shara et al. 2010; hereafter Y05, S10). Typical ejecta masses are in
the $10^{-5}$--$10^{-4}\, {\rm M_\odot}$ range; but Y05 and S10 showed
that nova eruptions on slowly-accreting, low-temperature WDs can eject
up to $\sim 2.2 \times 10^{-3}\, {\rm M_\odot}$ of gas.
%\footnote{Obtained for a $0.5\, {\rm M_\odot}$ WD with
%  surface temperature $3 \times 10^6\, {\rm K}$, accreting at a rate
%  $\dot{m} \simeq 5 \times 10^{-11}\, {\rm M_\odot\, yr^{-1}}$.}
Ejecta mass estimates in X-ray detected novae (Pietsch et al. 2007)
somewhat confirm theoretical expectations: ejecta masses of $\ge
10^{-3}\,{\rm M_\odot}$ are likely in at least 2 out of 18 novae.
%\footnote{Table~8 of
%  Pietsch et al. 2007 actually lists 4 such objects; but the 2 highest
%  estimates of the ejecta masses are unreliable because they involve
%  recurrent novae.}.

WD-hosting binaries are common in GCs (Sigurdsson \& Phinney 1995);
thus, novae should be frequent in GCs. This disagrees with the low
number of novae observed in GCs\footnote{We are aware of 5 GC novae: 2
  in Milky Way GCs, 2 in M31, 1 in M87; see Shara et al. (2004),
  Shafter \& Quimby (2007), Henze et al. (2009), Moore \& Bildsten
  (2011); a sixth nova candidate (in a M31 GC) was reported by Peacock
  et al. (2010).}; but the discrepancy might be explained by
observational biases.  We parametrize the nova rate in a GC as
$\Gamma_{\rm GC}=\gamma_{\rm GC} \Gamma_E M_{\rm GC}$, where $\Gamma_E
\sim 2.2\times10^{-10}\; {\rm yr^{-1}\, M_\odot^{-1}}$ is the nova
rate per unit stellar mass in elliptical galaxies (Della Valle et
al. 1994; Mannucci et al. 2005; Henze et al. 2009), $M_{\rm GC}$ is
the GC mass, and $\gamma_{\rm GC}$ is the enhancement of the nova rate
in GCs (compared to ellipticals). We expect $\gamma_{\rm GC} \gtrsim
1$ because of the over-abundance of WD-hosting binaries in GCs. On the
other hand, observations of the M31 and M87 GC systems can put upper
limits on nova occurrence rates: for M87, Shara et al. (2004) infer
$\Gamma_{\rm GC,M87}\sim 0.004\ {\rm yr^{-1}\, GC^{-1}}$, whereas for
M31 we have an upper limit $\Gamma_{\rm GC,M31}\lesssim 0.005\; {\rm
  yr^{-1}\, GC^{-1}}$ (Henze et al. 2009): assuming $M_{\rm
  GC}\sim10^6\,{\rm M_\odot}$, these rates imply $\gamma_{\rm
  GC}\lesssim 20$.

%, or an estimate of $\sim 0.002 \; {\rm yr^{-1}\, GC^{-1}}$
%by Hanze et al. (2009)\footnote{Such estimate is based upon the
%combination of a novae rate of $\sim 2.2\; {\rm yr^{-1}\,
%(10^{10}\,M_\odot)^{-1}}$ in elliptical galaxies (Della Valle et
%al. 1994; Mannucci et al. 2005) with a total mass of the M31 GC system
%of $5\times10^9\, {\rm M_\odot}$; the latter might be an overestimate
%by a factor $\sim 10$, since it implicitly assumes that the average
%mass of the $\sim 500$ known M31 GCs is $\sim 10^7\,{\rm M_\odot}$;
%however, this might be partially compensated by the fact that inside
%GCs the fraction of WD in a close binary is likely larger than in an
%elliptical galaxy.}.

\section{Modelling of line emission}
We use the photoionization code {\tt Cloudy} (version 08.00, see
Ferland et. al. 1998) to model the optical spectrum resulting from the
photoionization of the nova ejecta by the XRS.

%\subsection{Grid of models}
We represent the nova ejecta as a slab of material with thickness
$\Delta d$, number density $n$, filling factor $f_{\rm fill}$, and
metallicity $Z$, at a distance $d$ from the XRS, covering a fraction
$f_{\rm cov}$ of the solid angle around it. We note that $f_{\rm cov}$
determines two different geometrical configurations:
\begin{enumerate}
\item{}{if $f_{\rm cov}$ is large ($\ge 0.5$), the XRS must be {\it within}
  the nova ejecta. In such case, we neglect the offset
  between the XRS and the centre of the ejecta shell, so that the
  radius of the ejecta shell is $r_{\rm ej}\sim d$; then, the
  time elapsed after the nova explosion is $t_{\rm nova}\sim r_{\rm
    ej}/v_{\rm ej} \simeq 30\,{\rm yr}\ d_{17}\, v_{\rm ej,3}^{-1}$,
  where $d_{17}\equiv d/(10^{17}\,{\rm cm})$, and $v_{\rm ej,3}\equiv
  v_{\rm ej}/(10^3\,{\rm km\,s^{-1}})$.}
\item{}{if $f_{\rm cov}$ is small, we assume the XRS to be {\it
  outside} the ejecta, and estimate $r_{\rm ej}$ from simple
  geometry: $f_{\rm cov}\simeq(\pi r_{\rm ej}^2)/(4\pi d^2)$, so that
  $t_{\rm nova}\sim 2 \sqrt{f_{\rm cov}} (d/v_{\rm ej}) \simeq 6\,{\rm
  yr}\ (f_{\rm cov}/0.01)^{1/2}\, d_{17}\, v_{\rm ej,3}^{-1}$.}
\end{enumerate}

The XRS spectrum was taken from M07: we use a disk
blackbody spectrum with $kT=0.22\,{\rm keV}$ and 0.2--10 keV
luminosity $\sim 4.5\times10^{39}\, {\rm erg\, s^{-1}}$. We note that
M07 reported a large variation (a drop by a factor of $\sim 7$) in the
count rate during a single {\it XMM-Newton} observation. Such
variation is consistent with an increase in the intervening
column density, from $N_{\rm H} \simeq 1.67 \times 10^{20}\, {\rm
cm^{-2}}$ (consistent with the Galactic value) to $N_{\rm H}$
$\sim 3 \times 10^{21}\, {\rm cm^{-2}}$. Thus, we assume that the
intrinsic X-ray spectrum does not change; but we experiment with
different $N_{\rm H}$ values.

%%%%%%%%%%%%%%%%%%%%%%%%%%%%%%% TABLE 1%%%%%%%%%%%%%%%%%%%%%%%%%%%%%%%%%
\begin{table}
\begin{center}
\caption{Explored parameter space.} \leavevmode
\begin{tabular}[!h]{lc}
\hline
Parameter & Range of values\\
\hline
$\log{d}$ [cm]                      & 16--18.5, with 0.25 steps\\
$\log{\Delta d}$ [cm]               & 15.0, 15.5, 16, 16.5\\
$f_{\rm cov}$                       & 0.003, 0.01, 0.03, 0.1, 2/3 \\
$\log{n}$ [cm$^{-3}$]               & -2 to 10, with 0.1 steps\\
$f_{\rm fill}$                      & 2/3 \\
$Z$                                 & 0.14, 0.29, 0.50\\
\hline
$kT$ [keV]                          & 0.22\\
${L_{0.2-10}}^{(a)}$ [erg s$^{-1}$]   & $4.5 \times 10^{39}$ \\
${\log{N_{\rm H}}}$ [cm$^{-2}$] & 19, 20, 21, 21.5\\
\hline
${m_{ej}}$ [M$_\odot$]              & $\le 10^{-2}$\\
\noalign{\vspace{0.1cm}}
\hline
\end{tabular}
\end{center}
\footnotesize{Definitions: $d$ is the distance from the XRS to the WD
  that produced the nova eruption; $\Delta d$ is the thickness of the
  nova shell; $f_{\rm cov}$ is the covering factor of the nova shell
  (as seen from the XRS); $n$ is the number density of the nova shell
  ($n\equiv \rho/m_{\rm H}$, where $m_{\rm H}$ is the mass of the H
  atom); $f_{\rm fill}$ is the filling factor of the nova shell; $Z$
  is the mass fraction of metals; $kT$ is the temperature of the
  blackbody spectrum of the XRS; ${L_{0.2-10}}$ is the luminosity of
  the XRS in the band 0.2--10 keV; $N_{\rm H}$ is the column density
  that the radiation from the XRS must cross before reaching the nova
  shell; $m_{\rm ej}$ is the mass of the ejecta shell.\\
{\it Notes}: $^{(a)}$: derived assuming a distance $D=16$ Mpc.}
\end{table}
%%%%%%%%%%%%%%%%%%%%%%%%%%%%%%%%%%%%%%%%%%%%%%%%%%%%%%%%%%%%%%%%%%%%%%%%

Table 1 describes the extent of the explored parameter space. We
point out that:
\begin{enumerate}
\item{We use the {\tt Cloudy} pre-defined {\tt nova} abundance set. It
    represents typical nova ejecta\footnote{From Ferland \& Shields
    1978, based on observations of nova V1500 Cygni; solar abundances
    (from Grevesse \&\ Sauval 1998, Holweger 2001, Allende Prieto,
    Lambert \&\ Asplund 2001, 2002) were used for elements not listed
    in that paper.}, and corresponds to an O/H number ratio of
    $1.7\times10^{-2}$ ($\sim30$ times larger than in the solar
    mixture), and to a metal mass fraction $Z\simeq0.29$. We obtain
    mixtures with different metallicities by rescaling the metal
    abundances of this set by a factor 0.4 ($Z=0.14$) or 2.5
    ($Z=0.50$).}
\item{The slab mass is $m_{\rm ej} = 4\pi d^2\, \Delta d\,
  f_{\rm cov}\, f_{\rm fill}\, n\, m_{\rm H} \simeq 10^{-2}\, {\rm
    M_\odot}\ d_{17}^2 [\Delta d/(10^{16}\,{\rm cm})] f_{\rm cov}
  f_{\rm fill} [n/(10^4\,{\rm cm^{-3}})]$, where $m_{\rm
    H}=1.67\times10^{-24}\,{\rm g}$ is the mass of the H atom. Since
  we expect that $m_{\rm ej} \lesssim 2.2\times 10^{-3}\,{\rm
    M_\odot}$ (Y05, S10), ionization models were calculated only when
  $m_{\rm ej}\le 0.01\,{\rm M_\odot}$.}
\end{enumerate}

%%%%%%%%%%%%%%%%%%%%%%%%%%%%%%%%%%% FIGURE 1 %%%%%%%%%%%%%%%%%%%%%%%%%%%%%%%%%%
\begin{figure}
\center{{ \epsfig{figure=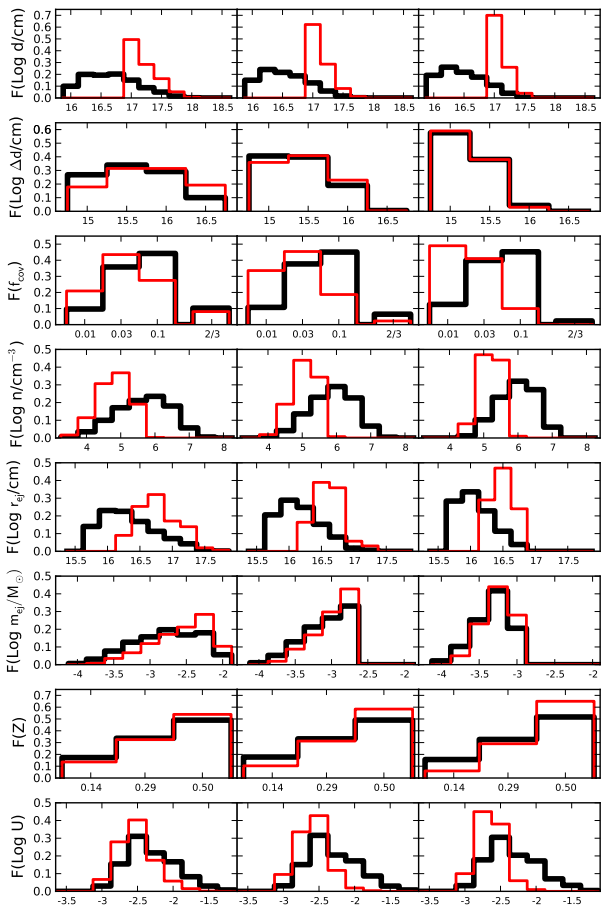,height=12cm} }}
\caption{\label{fig:fig1} Distribution of parameters for the viable
  models. Each row shows the distribution of one parameter (see the
  caption of Table~1; in addition, $U$ is the ionization parameter);
  columns correspond to different priors on the ejecta mass $m_{\rm
  ej}$ (left: $\le 10^{-2}\,{\rm M_\odot}$; centre: $\le 2.5 \times
  10^{-3}\,{\rm M_\odot}$; right: $\le 10^{-3}\,{\rm M_\odot}$). The
  thick lines refer to all the viable models within the $m_{\rm ej}$
  limit, whereas the thin (red on the web) lines further require the
  distance $d$ of the nova from the XRS to be $\ge 10^{17}\,{\rm
  cm}$. The distributions are normalized to the number of viable
  models satisfying the relevant prior (so, the thin line is often
  above the thick line even if it refers to a lower number of
  viable models).}
\end{figure}
%%%%%%%%%%%%%%%%%%%%%%%%%%%%%%%%%%%%%%%%%%%%%%%%%%%%%%%%%%%%%%%%%%%%%%%%%%%%%%%

\section{Comparison of models and data for RZ~2109}
Models were required to agree with two basic observational constraints
from the spectrum shown in Z08\footnote{The Z08 spectrum (taken in
December 2007) is quite similar to the other three spectra of RZ~2109
discussed in S11 (and taken in the following $\sim$1.3 years). Instead
the two spectra described by Z07 (taken in 2004--2005, i.e. in the
previous $\sim$3.9 years) appear quite different (FWHM
$\sim200$--$350\, {\rm km\, s^{-1}}$,
$F_{5007}\sim0.13\times10^{-16}\, {\rm erg\, s^{-1}\, cm^{-2}}$).  Z07
warn that uncertainties amount to a factor of $\gtrsim 2$, and the
discrepancy might be explained by the (very different) observational
setups.}
\begin{enumerate}
\item{}{the flux of the [OIII]$\lambda5007$ line, $F_{5007} \simeq
  4.4\times10^{-16}$. Z08 do not quote an error on this measurement:
  the equivalent width comparison performed by S11 suggests an
  uncertainty of 20--30 per cent; but the discrepancy with the Z07
  spectrum suggests a larger amount. Furthermore, we account for the
  approximations of the ionization models by increasing such
  uncertainty. Thus, we consider viable models with $F_{5007}$
  within a factor of 3 from the Z08 estimate.}
\item{}{The absence of any other emission lines in all the spectra: we
  exclude all the models where the luminosity in any optical
  line\footnote{We actually considered only H$\alpha$, H$\beta$,
  [OI]$\lambda6300$, [OII]$\lambda3727$,
  [NII]$\lambda\lambda6548,6583$, and [SII]$\lambda6716,6731$.}
  exceeds $0.1 L_{5007}$.}
\end{enumerate}

About 4 per cent of the models we calculated\footnote{We calculated a
total of $\sim 53000$ models, of whom 2144 are viable (655 if we
consider only models with $d \ge 10^{17}\, {\rm cm}$). If we require
$m_{\rm ej}\le 2.5\times 10^{-3}{\rm M_\odot}$ the number of viable
models is 1278 (262 if we further impose $d \ge 10^{17}\, {\rm cm}$),
and requiring $m_{\rm ej}\le 10^{-3}{\rm M_\odot}$ further decreases
this number to 652 (100 of whom have $d \ge 10^{17}\, {\rm cm}$).}
satisfy the above criteria.  In Fig.~1, we show the distribution of
the various parameters (including the ionization parameter, defined as
the number ratio of ionizing photons to H atoms, $U=Q(H)/(4\pi d^2 c
n)$, where $Q(H)$ is the number of ionizing photons - i.e. with energy
$\ge 13.6\, {\rm eV}$ - emitted by the XRS in the unit time) for these
viable models, and how these distributions are influenced by the
choice of priors about the models (e.g. the effects of using different
thresholds on the ejecta mass, or imposing a minimum for the nova-XRS
distance). Such distributions show us where a large amount of
parameter space is available.
%(or, more precisely, show us where little or none is available).
It can be seen that:

\begin{enumerate}
\item{it is almost impossible to have viable models if $d \gtrsim
  10^{17.5}\, {\rm cm} \simeq 0.1\, {\rm pc}$;}
\item{thin shells ($\Delta d \lesssim 10^{16}\, {\rm cm}$) are
favoured, especially when high-$m_{\rm ej}$ models are excluded;}
\item{models where the XRS is inside the nova shell (i.e. with $f_{\rm
cov}\ge 0.5$) are moderately unlikely, especially when high-$m_{\rm
ej}$ models are excluded;}
\item{shell densities are in the $10^4 \lesssim n/{\rm cm^{-3}}
\lesssim 10^7$ range; but models with high values of $d$ show a
narrower distribution, favouring the low-$n$ part;}
\item{$Z=0.5$ models are favoured; but it is possible to get
reasonable results if $Z=0.29$ or even $Z=0.14$.}
\item{The ionization parameter is mostly close to $\log(U) \sim -2.5$,
a fairly typical value for HII regions; this is somewhat remarkable,
as the ionizing spectrum is not produced by hot stars, and most of
the radiation energy is in the form of quite energetic (0.1--1 keV)
photons (rather than photons just above the Lyman limit).}
\end{enumerate}

\section{Discussion}

%%%%%%%%%%%%%%%%%%%%%%%%%%%%%%%%%%% FIGURE 2 %%%%%%%%%%%%%%%%%%%%%%%%%%%%%%%%%%
\begin{figure}
\center{{ \epsfig{figure=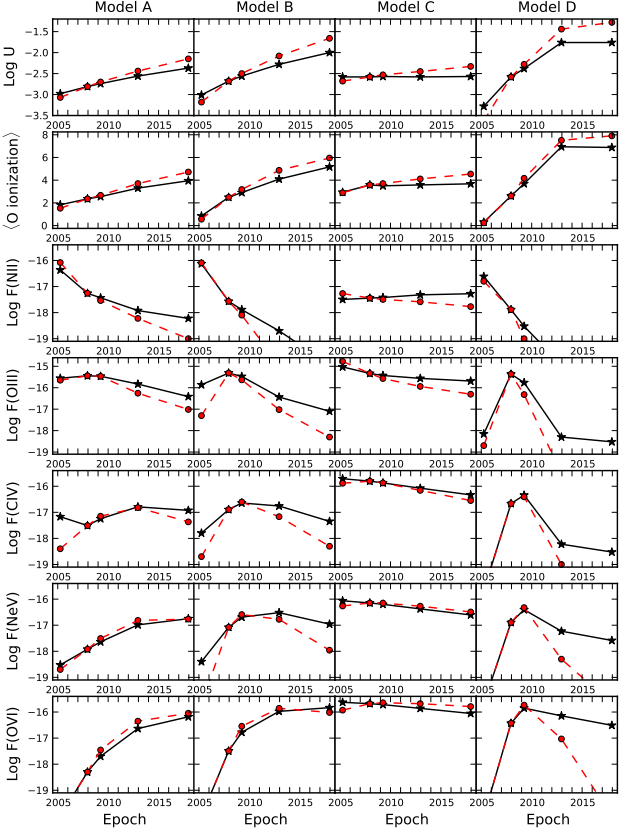,height=11cm} }}
\caption{\label{fig:fig2} Temporal evolution of model predictions for
  the four representative models described in the text; solid lines
  with star symbols are based on assumption (a) (i.e. that the shell
  thickness is constant), and dashed lines with round symbols (red on
  the web) are based on assumption (b) (i.e. that shell thickness is
  proportional to the ejecta radius). Each column refers to one of
  those models (model A to model D, from left to right), and all the
  plots have time on the x-axis. On the y-axis we show the ionization
  parameter $U$ (this is actually an assumption of the model, rather
  than a prediction; first row from the top), the average ionization
  stage of oxygen (i.e. $\sum_i{(i f_{{\rm O,}i})}$, where $f_{{\rm
  O,}i}$ is the fraction of O atoms that is ionized $i$ times; second
  row), and the flux at Earth (in ${\rm erg\, s^{-1}\, cm^{-2}}$) of
  five different lines or groups of lines ([NII]$\lambda6583$\AA,
  third row; [OIII]$\lambda$5007\AA, fourth row;
  CIV$\lambda\lambda1549,1551$\AA, fifth row; [NeV]$\lambda3426$\AA,
  sixth row; the complex of OVI lines between $\lambda1032$\AA\ and
  $\lambda1038$\AA, seventh row).}
\end{figure}
%%%%%%%%%%%%%%%%%%%%%%%%%%%%%%%%%%%%%%%%%%%%%%%%%%%%%%%%%%%%%%%%%%%%%%%%%%%%%%%

%%%%%%%%%%%%%%%%%%%%%%%%%%%%%%% TABLE 2%%%%%%%%%%%%%%%%%%%%%%%%%%%%%%%%%
\begin{table}
\begin{center}
\caption{Parameters of selected models.} \leavevmode
\begin{tabular}[]{lcccc}
\hline
Parameter & Model A & Model B & Model C & Model D\\ %A=1 B=2 C=6 D=8
\hline
$\log{d}$ [cm]                      & 17.25 & 17.00 & 16.75 & 16.50\\
$\log{\Delta d}$ [cm]               & 15.5  & 15.0  & 15.0  & 15.5\\
$f_{\rm cov}$                        & 0.03  & 0.03  & 2/3   & 0.10\\
$\log{n}$ [cm$^{-3}$]                & 4.7  & 5.5   & 5.0   & 5.5\\
$\log{U(13.6{\rm eV})}$             & -2.81 & -2.68 & -2.58 & -2.58\\
$Z$                                 & 0.50  & 0.29  & 0.50  & 0.29\\
${\log{n_{\rm H}}}$ [cm$^{-2}$] & 21   & 20    & 21.5  & 21.5\\
$m_{\rm ej}$ [10$^{-4}$ M$_\odot$]    & 10.6 & 6.7   & 14.9   & 7.0\\
$r_{\rm ej}$ [10$^{16}$ cm]          & 6.2  & 3.5    & 5.6   & 2.0\\
$t_{\rm nova}$ [yr]                   & 13.1 & 7.4   & 11.9  & 4.2\\

\noalign{\vspace{0.1cm}}
\hline
\end{tabular}
\end{center}
\footnotesize{Parameters not included in this list (e.g. $L_{0.2-10}$)
  are fixed at the values in Table~1; the values given here refer to
  the epoch of the Z08 spectrum (December 2007). The estimate of
  $t_{\rm nova}$ assumes $v_{\rm ej,3}=1.5$.}
\end{table}
%%%%%%%%%%%%%%%%%%%%%%%%%%%%%%%%%%%%%%%%%%%%%%%%%%%%%%%%%%%%%%%%%%%%%%%%

\subsection{Predictions}

[OIII] emission should evolve on a timescale $t_{\rm nova}$. Thus,
our scenario can be tested by monitoring the RZ~2109 spectrum.

We arbitrarily select four viable models (see Table~2) for a study of
the time evolution of line emission. This was done by assuming that
the nova shell keeps expanding with $v_{\rm ej,3}\simeq 1.5$
(cfr. Z08, S11), and properly scaling the relevant parameters ($f_{\rm
cov}$, $\Delta d$ and $n$ if XRS is outside the expanding shell; $d$,
$\Delta d$ and $n$ otherwise). As the evolution of the thickness
$\Delta d$ of the nova shell is unclear, we consider two different
cases: (a) $\Delta d=$constant (appropriate for small velocity
dispersions within the ejecta), and (b) $\Delta d \propto r_{\rm ej}$
(appropriate for a significant ejecta velocity dispersion).

We calculated the properties of each of these four models at five
times: three past epochs (April 2005, December 2007, and March 2009,
i.e. the times of the last Z07 observation, of the Z08 observation,
and of the last S11 observation, respectively), and two future ones
(December 2012 and December 2017, i.e. 5 and 10 years after the Z08
observation). The behaviour of the results are summarized in Fig.~2,
where we show the time evolution of the ionization parameter, of the
ionization state of Oxygen, and of the intensities of five lines (or
groups of lines). Such lines are representative of the main ionization
stages and of the most abundant heavy elements in the ejecta, but are
{\it not} always the most intense: an exhaustive list of possibly
detectable optical and UV lines can be found in the Appendix.

%% In Tables~3, 4, 5, and 6 we list the results for three past and two
%% future epochs: the epoch labeled 2007 is taken to coincide with the
%% Z08 observation, the epochs labeled 2005 and 2008 were chosen to
%% coincide with the second observation reported by Z07 (2.7 years before
%% the one in Z08) and the last observation described in S11 (1.3 years
%% after the one in Z08), respectively; the 2012 and 2017 epochs refer to
%% hypothetical observations 5 and 10 years after the one in Z08.

In 7 out of 8 models the ionization parameter increases with time, so
that heavy elements (such as oxygen) move to higher ionization
states. Therefore, high-ionization lines (such as the OVI lines around
$\lambda$1035\AA) tend to grow stronger, whereas low-ionization lines
(e.g. [NII]$\lambda\lambda$6548,6583\AA) tend to weaken with time, and
intermediate-ionization lines (e.g. the CIV doublet at
$\lambda\lambda$1549,1551\AA) grow to a peak, and then start to
decline.

The case (a) of model C is the main exception: since it assumes
that (i) the XRS is within the expanding shell, and (ii) the shell
thickness is constant, the density evolution in the shell ($n\propto
r_{\rm ej}^{-2}$) compensates the geometrical dilution of the ionizing
flux, resulting in a constant ionization parameter. As a consequence,
the ionization state of Oxygen (and of all the other heavy elements)
remains practically constant, and line fluxes tend to decrease very
slowly.

%%  moves towards
%% higher ionization, so that high-ionization lines (e.g. the OVI doublet
%% at $\lambda\lambda=1032,1038$\AA) grow steadily stronger, while
%% low-ionization ones (e.g. [OII] at $\lambda=3727$\AA) become weaker
%% (lines of intermediate ionization states - e.g. [OIII] - grow to a
%% peak and then start to decline).  The exception is the case (a) of
%% model C, where by construction the ionization parameter is constant
%% (the model assumes that the XRS is within the expanding shell, and the
%% $n\propto r_{\rm ej}^{-2}$ density evolution on the shell compensates
%% the geometrical dilution of the ionizing flux): such case exhibits the
%% slowest flux evolution, and very little evolution in the ionization
%% structure.  [WE COULD ADD A FIGURE SHOWING THE EVOLUTION OF EMISSION
%%   LINES AND OTHER PARAMETERS]

The four models were required to agree with the December 2007
observations, so their spectra for this epoch are dominated by the
[OIII] doublet, with $F_{5007}$ within 20 per cent from the Z08
value. In addition, [NeIII]$\lambda3869$ and [OIII]$\lambda4363$ might
be strong enough to be detectable: the latter was marginally detected
by Z08; as to [NeIII], we note that the Z08 spectrum is very noisy at
$\lambda\lesssim$4000\AA, and this line might be confused with other
features in that region. All the models (except model D - see next
subsection) are in quite satisfactory agreement with the 2009
observation, too: this is simply because the 1.3 years interval from
December 2007 to March 2009 is a relatively small fraction of $t_{\rm
nova}$.

Instead we find larger (and more interesting) differences if we look
at the April 2005 spectrum: models A and C predict that the [OIII]
emission was as strong (A) or stronger (C) than in 2007, and require
that the discrepancy among the fluxes reported by Z07 and Z08 derives
from observational factors only. Instead, models B and D provide a
different explanation: their quite short $t_{\rm nova}$ imply large
differences in $r_{\rm ej}$: in 2005 $r_{\rm ej}$ was only 2/3 (model
B) or 1/3 (model D) of its value in 2007; such a difference not only
reduces the amount of energy intercepted by the ejecta, but at the
same time lowers the ionization parameter, favouring low-ionization
states and moving the emission from [OIII] to [OII] (or even [OI])
lines.

The predictions for 2012 and 2017 offer an opportunity to test these
models, with model D predicting that all anomalous emission lines
should disappear before 2012, model B predicting large declines, and
models A and C predicting a slower, steady decline.  As we already
mentioned, such decline is mostly understood in terms of an increasing
ionization parameter, so that the O$^{++}$ abundance drops.

\subsubsection{Modelling issues}

It should be noted that all the above predictions suffer from a number
of approximations:
\begin{enumerate}
\item{The structure of a nova shell is not uniform. There are large
density fluctuations both at early times ($\lesssim 2$ years after the
eruption; see e.g. the images of Nova Cygni 1992 in Chochol et
al. 1997) and at late times ($\gtrsim 10$ years after the eruption;
see e.g. the images of T Pyx in Shara et al. 1997).}
\item{The residual nuclear burning on the WD responsible for the nova
eruption might produce a significant amount 
of UV/X-ray photons for a few years (see Henze et al. 2010): furthermore,
when $t_{\rm nova}\lesssim 2{\rm yr}$, $r_{\rm ej}\le 10^{16}{\rm cm} \ll d$,
and even moderate WD luminosities (say $L_{\rm X, WD} \sim 10^{38}\,
{\rm erg\, s^{-1}\, cm^{-2}}$) are enough for the WD emission to dominate
over the one from the XRS. Therefore, the predictions for model B
in 2005, for model D in 2007, and especially for model D in 2005 might
suffer from the fact that our models do not consider the WD emission.}
\item{The {\tt Cloudy} calculations assume steady state, but models
with a long ($\gtrsim$1 yr) recombination timescale would require
a time-dependent approach. For H recombination, $t_{\rm rec,H}\sim
(\alpha_{\rm A} n)^{-1} \simeq 0.8\, (n/10^5\,{\rm cm^{-3}})^{-1} {\rm
yr}$ (where $\alpha_{\rm A}\sim 4\times10^{-13}\,{\rm cm^3\, s^{-1}}$
is the H recombination coefficient), so that models with $n\lesssim
10^5\, {\rm cm^{-3}}$ might suffer from this problem; however, the
recombination timescales of highly ionized metals are shorter (by
factors of 2--10) than $t_{\rm rec,H}$, and predictions about their
lines should be reliable for $n\gtrsim 10^4\,{\rm cm^{-3}}$.}
\item{Our geometry is extremely simple, and might suffer
from assuming that the whole nova shell is at the same
distance from the XRS: this is reasonable if the XRS is within the
shell (model C), or if $r_{\rm ej}\ll d$, but there are a number
of cases (models A and B at the 2012 and 2017 epochs; model D at all
epochs except 2005) where $r_{\rm ej}\ge d/2$ (so that the ionization
parameter in the XRS-approaching part of the shell might be $\ge 9$
times higher than in the XRS-receding part)\footnote{Epochs 2012 and
2017 of model D are extreme cases, since we have $r_{\rm ej}> d$, and
the XRS is engulfed by the shell; we try to adapt to the new situation
by switching to a model with the XRS at the centre of the shell
($f_{\rm cov}=2/3$, $d=r_{\rm ej}$), but these two set of predictions should
be considered extremely uncertain.}.}
\item{The timing of the evolution depends upon the exact value of
$v_{\rm ej,3}$: the adopted 1.5 value is reasonable (S11 estimates
$v_{\rm ej,3}\simeq 1.3$ for the broad component of the line profile -
see the next subsection), but not unique: a lower(higher) choice of
$v_{\rm ej}$ would delay(fasten) the predicted evolution (for example,
if $v_{\rm ej}=1$, our predictions for December 2012 and December 2017
would apply to June 2015 and December 2022, respectively). This is
particularly relevant for Model D: the [OIII] flux if Fig.~2 and
Table~A4 evolves too rapidly (the 2007 flux agrees with the
observations, but both the 2005 and the 2009 fluxes are too low);
however, the model should not be discarded (yet) because these
discrepancies can still be reconciled using a value of $v_{\rm ej,3}$
($\simeq 1$) that is still in agreement with the S11 measurements.}
\end{enumerate}

Some of these issues might be resolved by adopting more complicated
modelling (e.g. assume a spectrum/temporal evolution for the WD
emission, use some density distribution, adopt a more realistic
geometry, etc.); however, a more detailed modelling, while providing
more reliable predictions, would also require us to introduce a number
of additional parameters. This is not only a problem in terms of
resources (such as computing power), but would further complicate the
interpretation of the results.

\subsection{Line profile}
S11 model the spectral profile of the [OIII]$\lambda\lambda4959,5007$
complex as the superposition of the identical profiles of the two
lines of the doublet, each one made of a central, relatively narrow
(FWHM$\sim 300\, {\rm km\, s^{-1}}$) gaussian peak, plus a much
broader (FWHM$\sim 1300\, {\rm km\, s^{-1}}$) `flat' component
(possibly the result of a bipolar outflow); notably, the latter
component is not really flat: the red side is about 1.4 times brighter
than the blue side. As noted by S11, this is consistent with a nova
shell between the XRS and the observer\footnote{S11 also note that a
  nova shell fails to explain why the line profile is made of 2
  components. We do not attempt to build a model, but it is quite
  likely that such a profile can be reproduced if the shell density is
  not uniform. Alternatively, the nova shell might be made of a slow
  ($v\sim 300\,{\rm km\,s^{-1}}$) isotropic component, plus a fast
  bipolar one ($v\sim 1300\,{\rm km\,s^{-1}}$). Nova observations
  (e.g. Iijima \& Naito 2011) and theoretical models (see e.g. the
  ejecta velocity plots in S10) support these hypotheses by
  finding/predicting complex profiles for nova emission lines.}.  If
this is the case, the flux difference among the two sides suggests
that in 2007-2008 $r_{\rm ej}$ was of the order of 10 per cent of the
distance between the nova and the XRS: this would favor models with
$f_{\rm cov}\sim$ 0.01. Furthermore, since $r_{\rm ej}$ increases with
time, the asymmetry between the red and blue sides should increase as
well (at least as long as the XRS is outside the nova shell).

If the nova lies between us and the XRS, the ejecta might even
intercept the line of sight to the XRS, increasing the value of
$N_{\rm H}$. We checked whether such an increase can explain the
fading of the XRS seen in the 2004 {\it XMM} observation (see M2007),
and in the subsequent {\it Swift} and {\it Chandra} observations (see
Maccarone et al. 2010). The effective X-ray extinction due to
isotropic nova ejecta is
\begin{eqnarray}
N_{\rm H,shell} & = g {m_{\rm ej} \over{4\pi r_{\rm ej}^2 m_{\rm H}}}
{Z\over Z_{\odot}} \simeq \nonumber\\ & \simeq 4.2\times10^{22} {\rm
  cm^{-2}} \left({g\over3}\right) \left({m_{\rm ej}\over{10^{-3}\,{\rm
      M_\odot}}}\right) \left({r_{\rm ej}\over{10^{16}\,{\rm
      cm}}}\right)^{-2} \left({Z\over{0.29}}\right)
\end{eqnarray}
where $g$ is a geometrical factor accounting for the number of times
that the photons cross the shell (1 if the XRS is within the shell; 2
if the XRS is outside the shell), and for the inclination of the
photon direction with respect to the normal to the shell (note that
$g$ can be quite larger than 2 if the line of sight to the XRS is
almost tangential to the ejecta shell). Maccarone et al (2010) suggest
that the weakness of the XRS in RZ~2109 during the {\it Chandra} 2010
observation might be explained by a large increase in the effective
$N_{\rm H}$ in front of the XRS. We used the web version of W3PIMMS
({\tt http://heasarc.nasa.gov/Tools/w3pimms.html}) to estimate that
$N_{\rm H}\gtrsim 3\times10^{22}\,{\rm cm^{-2}}$ is necessary to
explain the {\it Chandra} 2010 observations. This is marginally
consistent with our scenario: if we set $g=6$, $Z=0.50$, $m_{\rm
ej}=2.2\times10^{-3}\,{\rm M_\odot}$, we can obtain such a high
$N_{\rm H}$ for $r_{\rm ej}\lesssim3\times 10^{16}\,{\rm cm}$. In
turn, this implies that the nova eruption occurred in 2000--2004 (for
$1 \lesssim v_{\rm ej,3} \lesssim 1.5$ ), which might be consistent
with the timing of the Z07 observations (2004--2005). However, the
available parameter space is very small: a slight change in the choice
of $g$, $Z$ and $m_{\rm ej}$ can reduce the upper limit on $r_{ej}$,
moving the nova occurrence to an epoch {\it after} the first Z07
observation. Therefore, we deem quite unlikely that the weakening of
the XRS is (entirely) due to the additional absorption from the nova
shell; however, if this is the case, the XRS should re-brighten within
a few years.

\subsection{Probability}
We estimate the probability of having a nova shell with properties
suitable to explain the RZ~2109 observations within a distance
$\tilde{d}$ from the XRS as
\begin{eqnarray}
p(<\tilde{d}) & = \Gamma_E \gamma_{\rm GC} (N_{\rm GC}
    \langle m_{\rm GC} \rangle)\, 
    \left({{{4\pi}\over3} \tilde{d}^3}\right)
	 {{n_{\rm *,c}}\over{N_{\rm GC}}} \,
    \tilde{f} \, t_{\rm nova} \sim \nonumber\\
  &\sim 5\times 10^{-9} \,
    \gamma_{\rm GC} \,
    {{\langle m_{\rm GC} \rangle}\over{0.5 {\rm M_\odot}}} \,
    \left({{\tilde d}\over{0.1 {\rm pc}}}\right)^3 \,
    {{n_{\rm *,c}}\over{10^5 {\rm pc^{-3}}}} \,
    {{\tilde{f}}\over{0.01}} \,
    {{r_{\rm ej}}\over{0.01{\rm pc}}} \,
    v_{\rm ej,3}^{-1},
\end{eqnarray}
where $\Gamma_E$ and $\gamma_{\rm GC}$ were introduced in Sec. 2,
$N_{\rm GC}$ is the number of stars in a GC, $\langle m_{\rm GC}
\rangle$ is the average stellar mass in a GC, $n_{\rm *,c}$ is the
number density of stars at the centre of a GC ($10^4$--$10^6 \, {\rm
pc^{-3}}$; see e.g. Pryor \& Meylan 1993), $\tilde{f}$ is the fraction
of novae whose ejecta can produce the observed line emission (a highly
uncertain quantity; since relatively massive ejecta are needed, we use
0.01 as a reference), and $t_{\rm nova}\sim 10 \, {\rm yr} [r_{\rm
ej}/(0.01\,{\rm pc})] v_{\rm ej,3}^{-1}$ acts as the visibility
timescale for the [OIII] line emission.

Such a probability is of the same order of magnitude as that predicted
in other scenarios such as the disruption of a WD by an IMBH (Clausen
\& Eracleous 2011 quote a disruption rate of $\sim 10^{-8}\,{\rm
  yr^{-1}\,GC^{-1}}$ from Sigurdsson \& Rees 1997), and it must be
noted that the large uncertainty about several parameters in the above
formula might lead to a substantial increase of $p(<\tilde{d})$.

We note that, if the XRS is outside the nova shell, a further factor
$\Omega/4\pi$ accounting for the possible beaming of the XRS radiation
should be introduced into eq. (2): therefore, an observational
confirmation of this scenario might disfavour the models of the XRS
(e.g. King 2011) that assume a beamed emission\footnote{Beamed XRS
models face problems even when the XRS is within the shell: the
emission from gas outside the beam is likely negligible, so that the
fluxes of {\tt Cloudy} models assuming $f_{\rm cov}=2/3$ (appropriate
for this geometry) should be multiplied by a factor of $\le
(\Omega/4\pi)/(2/3)$: if $\Omega/4\pi \lesssim 0.1$, no viable model
of this kind would survive.}.

\section{Summary and conclusions}
We used the photoionization code {\tt Cloudy} to check whether the
peculiar emission spectra from XRS-hosting GCs (especially RZ~2109)
can be explained as the result of the photoionization of a nova shell
by the radiation emitted by the XRS. We found that this is definitely
possible, but requires that the nova ejecta are relatively massive
($\gtrsim 0.5\times10^{-3}\, {\rm M_\odot}$), and that the XRS-nova
distance is quite small ($\lesssim 0.1\, {\rm pc}$).

The combination of these two requirements is quite unlikely; on the
other hand, it can explain observations in a straightforward manner
(although the line profile needs further investigation). While not
entirely satisfactory, this compares well with the two other proposed
scenarios for RZ~2109: (i) the disruption of a WD by an IMBH (an event
as unlikely as the one we discuss) explains the line profile and the
absence of H lines (I10, Clausen \& Eracleous 2011), but appears to be
unable to produce the high [OIII] luminosity (Porter 2010, S11); (ii)
the ionization of material ejected from the XRS itself, while not
suffering from the ``unlikely coincidence'' problem, does not explain
either the absence of H emission lines, nor some properties of the
line profile (asymmetry in the flat component, origin of the gaussian
component).

Finally, we predict the future evolution of the line emission in the
XRS+nova scenario, by looking at both optical and UV lines. We find
that in most cases the nova shell tends to move towards a higher
ionization state, so that the line flux should move from
moderate-ionization lines such as [OIII]$\lambda5007$\AA\ to
high-ionization lines (e.g. the OVI lines in the 1032--1038\AA\
wavelength range). Such predictions might provide a crucial test for
discriminating among the various hypotheses described above.

\section*{Acknowledgments}
We thank the anonymous referee for a careful reading of the
manuscript, and for several very valuable suggestions (e.g. about UV
lines), and thank A.~Wolter, L.~Zampieri, A.~Bressan, D.~Fiacconi,
S. Hachinger, R. Porter, and T. Maccarone for useful discussions and
comments.  This research has made use of NASA's Astrophysics Data
System Bibliographic Services, and of tools provided by the High
Energy Astrophysics Science Archive Research Center (HEASARC; which is
provided by NASA's Goddard Space Flight Center).

\appendix

\section{Full predictions - Tables}

In this appendix, we list the detailed predictions about line fluxes
at Earth, and about the properties of the nova ejecta (e.g. the
ionization state of C, N, O, and Ne) from the four models described in
the text, evolved to five past or future epochs: April 2005, December
2007, March 2009 (chosen to coincide with the times of past optical
observations), December 2012, and December 2017.

We report all the emission lines in the UV-optical range (913\AA$\le
\lambda \le$10000\AA) reaching a flux $\ge 5\times10^{-18}\,{\rm
erg\,s^{-1}\,cm^{-2}}$ in at least one of the five considered epoch,
plus Lyman$\alpha$ and H$\alpha$.  For well-separated multiplets we
list only the strongest line (for example, in the case of the
[OIII]$\lambda\lambda$4959,5007\AA\ doublet we only list
[OIII]$\lambda$5007\AA), whereas close multiplets are listed as a
single line (for example the [OII]$\lambda\lambda$3726,3729\AA\ doublet
is listed as [OII]$\lambda$3727\AA).  All the fluxes are in units of
$10^{-16}\, {\rm erg\, s^{-1}\, cm^{-2}}$, and were obtained assuming
a distance $D=16\,{\rm Mpc}$. No reddening correction was applied.

All the data columns - except the one referring to the Z08
observation, which is taken as a reference point - report two set of
data: the first was obtained assuming that the shell thickness $\Delta
d$ remains constant at the value used for December 2007 (implying that
the density $n$ scales as $r_{\rm ej}^{-2}$;assumption a), whereas the
second was obtained assuming that $\Delta d$ increases linearly with
time (i.e. that $\Delta d/r_{\rm ej}$ remains constant at the value
estimated for December 2007; in such case the density $n$
scales as $r_{\rm ej}^{-3}$; assumption b). The two cases are
generally separated by a ``/'', but in the case of ionization states
it was necessary to use two separate lines.

%%%%%%%%%%%%%%%%%%%%%%%%%%%%%%% TABLE A1%%%%%%%%%%%%%%%%%%%%%%%%%%%%%%%%%
\begin{table*}
\begin{center}
\caption{Evolution of emission line fluxes and other physical
properties for model A.} \leavevmode
\begin{tabular}[]{lccccc}
\hline
Line & April 2005 & December 2007 & March 2009 & December 2012 & December 2017\\
\hline
\noalign{\bf Model A - Spectral lines}\\ %Model 1 in lines_evol.txt
%OIV$\lambda922.7$        &
%0/0         & 0     & 0.001/0.001 & 0.002/0.002 & 0.002/0.002\\
%NIV$\lambda923.2$        &
%0.002/0.002 & 0.004 & 0.004/0.004 & 0.003/0.002 & 0.002/0.001\\
%CIII$\lambda977$        &
%0.001/0.001 & 0.004 & 0.007/0.010 & 0.022/0.017 & 0.015/0.005\\
NIII$\lambda991$        &
0.020/0.014 & 0.043 & 0.063/0.074 & 0.129/0.094 & 0.077/0.026\\
OVI$\lambda1032$--$1038$   &
0/0         & 0.005 & 0.020/0.035 & 0.232/0.446 & 0.654/0.900\\
%NII$\lambda1085$        &
%0.015/0.012 & 0.012 & 0.009/0.008 & 0.004/0.002 & 0.002/0.001\\
%OVI$\lambda1125$         &
%0/0         & 0     & 0.001/0.001 & 0.002/0.003 & 0.004/0.004\\
%NeV$\lambda1141$        &
%0/0         & 0     & 0/0.001     & 0.005/0.010 & 0.014/0.020\\
%OV$\lambda1211$         &
%0/0         & 0.005 & 0.012/0.019 & 0.096/0.171 & 0.227/0.249\\
Lyman $\alpha$ $\lambda 1216$ &
0.168/0.182 & 0.103 & 0.081/0.072 & 0.045/0.033 & 0.029/0.018\\
OV]$\lambda1218$         &
0/0         & 0.004 & 0.011/0.016 & 0.076/0.131 & 0.170/0.178\\
NV$\lambda1239,1243$    &
0.005/0.002 & 0.055 & 0.131/0.189 & 0.667/0.878 & 0.977/0.704\\
%CII$\lambda1335$        &
%0.005/0.005 & 0.003 & 0.002/0.002 & 0.001/0.001 & 0.001/0\\
%OV$\lambda1371$        &
%0/0         & 0     & 0.001/0.001 & 0.002/0.002 & 0.003/0.002\\
OIV]$\lambda1402$        &
0.004/0.002 & 0.038 & 0.090/0.130 & 0.460/0.548 & 0.516/0.279\\
%NIV$\lambda1476$        &
%0.005/0.011 & 0.002 & 0.001/0.001 & 0.001/0     & 0/0        \\
NIV]$\lambda1486$       &
0.012/0.006 & 0.081 & 0.172/0.235 & 0.714/0.770 & 0.647/0.267\\
CIV$\lambda1549,1551$   &
0.068/0.004 & 0.031 & 0.057/0.071 & 0.162/0.151 & 0.119/0.043\\
OIII]$\lambda1661,1666$  &
0.030/0.020 & 0.092 & 0.135/0.147 & 0.168/0.088 & 0.073/0.025\\
NIII$\lambda1750$       &
0.067/0.046 & 0.196 & 0.286/0.314 & 0.383/0.203 & 0.146/0.037\\
%$[$NeIII]$\lambda1815$        &
%0/0         & 0     & 0.002/0.002 & 0.002/0.001 & 0.001/0\\
CIII]$\lambda1907,1909$ &
0.026/0.022 & 0.059 & 0.078/0.082 & 0.092/0.049 & 0.033/0.007\\
%NIII$\lambda2064$        &
%0.001/0.001 & 0     & 0.002/0.002 & 0.001/0.001 & 0.001/0\\
%NII]$\lambda2141$        &
%0.013/0.026 & 0.006 & 0.006/0.005 & 0.003/0.002 & 0.002/0.001\\
%$[$OIII]$\lambda2321$       &
%0.007/0.005 & 0.014 & 0.016/0.016 & 0.013/0.006 & 0.004/0.001\\
%CII]$\lambda2324-2329$   &
%0.005/0.009 & 0.004 & 0.003/0.003 & 0.002/0.001 & 0.001/0\\
$[$NeIV]$\lambda2424$       &
0.007/0.003 & 0.041 & 0.085/0.114 & 0.326/0.348 & 0.310/0.156\\
$[$OII]$\lambda2471$        &
0.050/0.100 & 0.014 & 0.011/0.008 & 0.004/0.002 & 0.002/0.001\\
%MgII$\lambda2796,2803$  &
%0.007/0.012 & 0.004 & 0.003/0.002 & 0.001/0     & 0/0        \\
%$[$NeV]$\lambda3346$        &
%0.001/0.001 & 0.004 & 0.008/0.011 & 0.038/0.056 & 0.065/0.062\\
$[$NeV]$\lambda3426$        &
0.003/0.002 & 0.012 & 0.023/0.031 & 0.103/0.153 & 0.178/0.171\\
$[$OII]$\lambda 3727$       &
0.055/0.098 & 0.012 & 0.010/0.008 & 0.005/0.002 & 0.003/0.001\\
$[$NeIII]$\lambda 3869$     &
0.166/0.148 & 0.232 & 0.242/0.229 & 0.139/0.058 & 0.041/0.010\\
%FeV$\lambda3892$        &
%0.002/0.002 & 0.004 & 0.006/0.006 & 0.007/0.004 & 0.003/0.001\\
%FeII$\lambda4300$ &
%0.001/0.004 & 0     & 0/0         & 0/0         & 0/0\\
$[$OIII]$\lambda 4363$      &
0.030/0.023 & 0.058 & 0.071/0.071 & 0.056/0.025 & 0.019/0.006\\
%H$\beta$ $\lambda 4861$&
%0.005/0.005 & 0.003 & 0.002/0.002 & 0.001/0.001 & 0.001/0    \\
$[$OIII]$\lambda 5007$      &
2.77/2.24   & 3.6   & 3.43/3.14   & 1.48/0.555  & 0.385/0.097\\
$[$NII]$\lambda 5755$       &
0.023/0.053 & 0.005 & 0.004/0.003 & 0.001/0.001 & 0.001/0    \\
%$[$OI]$\lambda 6300$       &
%0.001/0.012 & 0     & 0/0         & 0/0         & 0/0\\
H$\alpha$ $\lambda 6563$&
0.014/0.015 & 0.008 & 0.006/0.004 & 0.003/0.002 & 0.002/0.001\\
$[$NII]$\lambda 6583$       &
0.433/0.893 & 0.055 & 0.037/0.029 & 0.012/0.006 & 0.006/0.001\\
$[$OII]$\lambda 7325$       &
0.068/0.140 & 0.019 & 0.015/0.012 & 0.006/0.003 & 0.003/0.001\\
%SIII$\lambda9532$       &
%0.012/0.015 & 0.004 & 0.002/0.002 & 0/0         & 0/0    \\
\noalign{\vspace{0.1cm}}
\noalign{\bf Model A - Model parameters and ionization conditions}\\
$\log{n}$ [cm$^{-3}$]
& 4.87/4.96 & 4.70 & 4.63/4.59 & 4.45/4.33 & 4.26/4.04\\
$t_{\rm nova}\,{\rm [yr]}$
& 10.4 & 13.1 & 14.4 & 18.1 & 23.1\\ 
$r_{\rm ej}\,{\rm [10^{16} cm]}$
& 4.9  & 6.2  & 6.8  & 8.6  & 10.9\\ 
$f_{\rm cov}$
& 0.020 & 0.030 & 0.0354 & 0.533 & 0.0831\\
$\log{U(13.6{\rm eV})}$
& -2.98/-3.07 & -2.81 & -2.74/-2.70  & -2.56/-2.44 & -2.37/-2.15\\
C(I$\rightarrow$VII) &
%0,11.2,68.5,6.5,13.7  & 0.2,31.1,55.6,4.6,8.5 & %m27a, m27b
%0,2.8,55.2,11.5,30.6    & % 0
%0,1.9,44.8,13.1,40.3    & 0,1.5,40.1,13.4,45.1  & %p13a, p13b
%0,0.5,17.3,12.1,70.1    & 0,0.2,8.0,9.3,82.4    & %p50a, p50b
%0,0.3,5.1,6.4,88.2      & 0,0.1,1.3,2.4,96.2      %p100a, p100b
-,11,69,6,13,1,-    & %m27a
-,3,55,11,27,4,0    & %0
-,2,45,13,33,7,0    & %p13a
-,0,17,12,45,23,2    & %p50a
-,0,5,6,40,40,9        %p100a
\\
& 0,31,56,5,8,0,- & %m27b
&                         %0
-,1,40,13,36,9,1    & %p13b
-,0,8,9,44,33,5      & %p50b
-,0,1,2,26,50,21        %p100b
\\
N(I$\rightarrow$VIII) &
%0.1,25.2,63.1,6.8,4.7 & 2.9,47.8,42.2,4.5,2.7 & %m27a, m27b
%0,1.8,70.5,15.3,12.5    & %0
%0,1.1,60.8,19.9,18.2    & 0,0.8,55.7,22.0,21.5  & %p13a, p13b
%0,0.3,24.5,30.4,44.8    & 0,0.2,10.8,27.7,61.3  & %p50a, p50b
%0,0.2,7.3,20.9,71.7     & 0,0.1,1.9,9.1,88.9      %p100a, p100b
0,25,63,7,4,1,-,- & %m27a
-,2,70,15,9,4,0,-  & %0
-,1,61,20,11,7,1,-  & %p13a
-,0,24,30,15,25,5,0  & %p50a
-,0,7,21,14,40,16,1     %p100a
\\
& 3,48,42,4,2,0,-,- & %m27b
&                         %0
-,1,56,22,12,9,1,-    & %p13b
-,0,11,28,15,35,10,1    & %p50b
-,0,2,9,9,43,32,5        %p100b
\\
O(I$\rightarrow$IX) &
%0.2,30.8,57.5,9.4,2.1 & 2.5,51.7,38.8,5.6,1.2 & %m27a, m27b
%0,1.5,69.0,23.1,6.3     & %0
%0,0.9,58.3,31.1,9.6     & 0,0.7,52.7,34.8,11.8  & %p13a, p13b
%0,0.2,19.1,52.7,27.9    & 0,0.1,7.8,50.4,41.7   & %p50a, p50b
%0,0.1,6.0,41.2,52.7     & 0,0,2.1,21.8,76.1       %p100a, p100b
0,31,58,9,2,0,0,-,-   & %m27a
-,2,69,23,4,2,0,-,-     & %0
-,1,58,31,6,3,1,-,-     & %p13a
-,0,19,53,13,10,5,0,-    & %p50a
-,0,6,41,20,15,15,2,0       %p100a
\\
& 3,52,39,6,1,0,-,-,- & %m27b
&                         %0
-,1,53,35,7,4,1,0,-    & %p13b
-,0,8,50,18,13,10,1,-     & %p50b
-,-,2,22,20,17,30,8,0         %p100b
\\
Ne(I$\rightarrow$XI) &
%& 0.1,24.6,68.5,5.5,1.3 & 1.8,39.2,55.1,3.2,0.7 & %m27a, m27b
%0,1.5,79.4,15.5,3.7     & %0
%0,0.8,70.3,22.9,5.9     & 0,0.6,64.7,27.0,7.6   & %p13a, p13b
%0,0.2,28.6,49.3,21.9    & 0,0.1,12.9,50.2,36.8  & %p50a, p50b
%0,0.1,9.8,43.7,46.5     & 0,0,3.2,25.6,71.2       %p100a, p100b
0,25,69,5,1,0,-,-,-,-,-   & %m27a
-,1,79,15,3,0,-,-,-,-,-     & %0
-,1,70,23,5,1,-,-,-,-,-     & %p13a
-,0,29,49,19,3,0,-,-,-,-    & %p50a
-,0,10,44,36,9,1,0,0,-,-       %p100a
\\
& 2,39,55,3,1,-,-,-,-,-,- & %m27b
&                         %0
-,1,65,27,7,1,0,-,-,-,-     & %p13b
-,0,13,50,30,6,1,0,-,-,-    & %p50b
-,-,3,26,47,18,4,1,1,0,-      %p100b
\\
%S(I,II,III,IV,$>$V)
%& 0,4.9,51.3,26.7,17.2  & 0,16.8,54.2,18.7,10.3 & %m27a, m27b
%0,0.8,18.6,36.6,44.0    & %0
%0,0.4,10.3,31.2,58.1    & 0,0.3,7.8,27.6,64.3   & %p13a, p13b
%0,0,1.3,9.5,89.2        & 0,0,0.4,3.6,95.9      & %p50a, p50b
%0,0,0.3,1.7,98.0        & 0,0,0,0.2,99.8          %p100a, p100b
%\\
\noalign{\vspace{0.1cm}}
\hline

\end{tabular}
\end{center}
\footnotesize{Fluxes are in units of $10^{-16}\,{\rm
    erg\,s^{-1}\,cm^{-2}}$ and assume D=16 Mpc (no
    reddening). Wavelengths are in \AA, and in case of multiplets only
    the strongest component(s) are listed. The ionization states of
    C,N,O, and Ne are described by listing the percentages of each
    ionic state with respect to the total (with ``0'' indicating a
    percentage between 0.05\% and 0.5\%, and ``-'' indicating a
    percentage below 0.05\%). See text for more details}
\end{table*}
%%%%%%%%%%%%%%%%%%%%%%%%%%%%%%%%%%%%%%%%%%%%%%%%%%%%%%%%%%%%%%%%%%%%%%%%

%%%%%%%%%%%%%%%%%%%%%%%%%%%%%%% TABLE A2%%%%%%%%%%%%%%%%%%%%%%%%%%%%%%%%%
\begin{table*}
\begin{center}
\caption{Evolution of emission-line fluxes and other physical
properties for model B.} \leavevmode
\begin{tabular}[]{lccccc}
\hline
Line & April 2005 & December 2007 & March 2009 & December 2012 & December 2017\\
\hline
\noalign{\bf Model B - Spectral lines}\\ %Model 1 in lines_evol.txt
%OIV$\lambda922.7$        &
%0/0         & 0.002 & 0.004/0.004 & 0.005/0.003 & 0.002/0\\
%NIV$\lambda923.2$        &
%0/0         & 0.002 & 0.001/0.005 & 0.004/0.002 & 0/0\\
%CIII$\lambda977$        &
%0/0         & 0.016 & 0.029/0.030 & 0.016/0.005 & 0.004/0\\
NIII$\lambda991$        &
0.010/0.002 & 0.132 & 0.216/0.220 & 0.115/0.042 & 0.038/0.005\\
OVI$\lambda1032$--$1038$   &
0/0         & 0.032 & 0.167/0.285 & 1.06/1.39   & 1.47/0.961\\
%NII$\lambda1085$        &
%0.015/0.013 & 0.018 & 0.010/0.007 & 0.002/0.001 & 0/0\\
%OVI$\lambda1125$        &
%0/0         & 0     & 0.002/0.002 & 0.006/0.007 & 0.006/0.003\\
%NeV$\lambda1141$        &
%0/0         & 0.002 & 0.007/0.011 & 0.023/0.017 & 0.013/0.002\\
%OV$\lambda1211$         &
%0/0         & 0.017 & 0.072/0.115 & 0.323/0.298 & 0.258/0.095\\
Lyman $\alpha$ $\lambda 1216$ &
0.523/0.872 & 0.342 & 0.245/0.208 & 0.136/0.094 & 0.095/0.057\\
OV]$\lambda1218$         &
0/0         & 0.034 & 0.118/0.173 & 0.371/0.287 & 0.234/0.073\\
NV$\lambda1239,1243$    &
0.009/0.002 & 0.252 & 0.744/1.01  & 1.76/1.30   & 1.09/0.301\\
%CII$\lambda1335$        &
%0.009/0.013 & 0.004 & 0.002/0.002 & 0.001/0     & 0/0\\
%OV$\lambda1371$        &
%0/0         & 0.001 & 0.002/0.003 & 0.004/0.004 & 0.003/0.001\\
OIV]$\lambda1402$        &
0.007/0.001 & 0.181 & 0.495/0.637 & 0.607/0.268 & 0.200/0.030\\
%NIV$\lambda1476$        &
%0.020/0.026 & 0.001 & 0.001/0.001 & 0/0         & 0/0\\
NIV]$\lambda1486$        &
0.020/0.002 & 0.341 & 0.850/1.07  & 0.982/0.420 & 0.303/0.040\\
CIV$\lambda1549,1551$   &
0.016/0.002 & 0.127 & 0.228/0.250 & 0.175/0.068 & 0.045/0.005\\
OIII]$\lambda1661,1666$  &
0.067/0.003 & 0.306 & 0.347/0.285 & 0.077/0.024 & 0.022/0.002\\
NIII$\lambda1750$       &
0.140/0.006 & 0.655 & 0.809/0.710 & 0.221/0.065 & 0.056/0.004\\
%$[$NeIII]$\lambda1815$        &
%0/0         & 0.003 & 0.003/0.002 & 0/0         & 0/0\\
CIII]$\lambda1907,1909$ &
0.074/0.053 & 0.144 & 0.151/0.128 & 0.035/0.009 & 0.006/0\\
%NIII$\lambda2064$        &
%0/0         & 0.003 & 0.002/0.002 & 0.001/0     & 0/0\\
NII]$\lambda2141$        &
0.205/0.316 & 0.014 & 0.009/0.006 & 0.003/0.001 & 0/0\\
%$[$OIII]$\lambda2321$       &
%0.017/0.002 & 0.048 & 0.041/0.030 & 0.006/0.001 & 0.001/0\\
CII]$\lambda2324-2329$   &
0.067/0.114 & 0.007 & 0.004/0.003 & 0.001/0     & 0/0\\
$[$NeIV]$\lambda2424$       &
0.005/0     & 0.087 & 0.180/0.206 & 0.132/0.046 & 0.028/0.002\\
$[$OII]$\lambda2471$        &
0.546/0.798 & 0.024 & 0.011/0.006 & 0.002/0     & 0/0\\
%MgII$\lambda2796,2803$  &
%0.032/0.048 & 0.002 & 0/0         & 0/0         & 0/0\\
%$[$NeV]$\lambda3346$        &
%0.001/0     & 0.030 & 0.074/0.094 & 0.110/0.062 & 0.040/0.004\\
$[$NeV]$\lambda3426$        &
0.004/0     & 0.083 & 0.204/0.256 & 0.303/0.168 & 0.110/0.011\\
$[$OII]$\lambda 3727$       &
0.070/0.076 & 0.004 & 0.002/0.001 & 0.001/0     & 0/0\\
$[$NeIII]$\lambda 3869$     &
0.316/0.305 & 0.358 & 0.231/0.156 & 0.017/0.003 & 0.002/0\\
%FeV$\lambda3892$        &
%0/0         & 0.009 & 0.008/0.007 & 0.001/0     & 0/0\\
%FeII$\lambda4300$ &
%0.090/0.181  & 0    & 0/0         & 0/0         & 0/0\\
$[$OIII]$\lambda 4363$      &
0.071/0.003 & 0.206 & 0.176/0.129 & 0.024/0.006 & 0.005/0\\
%H$\beta$ $\lambda 4861$&
%0.009/0.009 & 0.009 & 0.006/0.005 & 0.003/0.002 & 0.001/0.001\\
$[$OIII]$\lambda 5007$      &
1.37/0.050  & 4.84  & 3.36/2.33   & 0.368/0.095 & 0.080/0.005\\
$[$NII]$\lambda 5755$       &
0.333/0.500 & 0.013 & 0.006/0.004 & 0.001/0     & 0/0\\
$[$OI]$\lambda 6300$       &
0.336/0.581 & 0     & 0/0         & 0/0         & 0/0\\
H$\alpha$ $\lambda 6563$&
0.027/0.030 & 0.025 & 0.017/0.014 & 0.008/0.005 & 0.004/0.002\\
$[$NII]$\lambda 6583$       &
0.748/0.812 & 0.027 & 0.013/0.008 & 0.004/0.001 & 0.001/0\\
$[$OII]$\lambda 7325$       &
0.726/1.06  & 0.033 & 0.015/0.009 & 0.002/0     & 0/0\\
%SIII$\lambda9532$       &
%0.015/0.020 & 0.001 & 0/0         & 0/0         & 0/0\\
\noalign{\vspace{0.1cm}}
\noalign{\bf Model B - Model parameters and ionization conditions}\\
$\log{n}$ [cm$^{-3}$]
& 5.83/6.00 & 5.50 & 5.38/5.32 & 5.10/4.90 & 4.82/4.48\\
$t_{\rm nova}\,{\rm [yr]}$
& 4.7 & 7.4 & 8.7 & 12.4 & 17.4\\ 
$r_{\rm ej}\,{\rm [10^{16} cm]}$
& 2.2 & 3.5 & 4.1 & 5.9  & 8.2\\ 
$f_{\rm cov}$
& 0.0139 & 0.030 & 0.0399 & 0.0759 & 0.1429\\
$\log{U(13.6{\rm eV})}$
& -3.01/-3.18 & -2.68 & -2.56/-2.50  & -2.28/-2.08 & -2.00/-1.66\\
C(I$\rightarrow$VII) &
1,60,33,2,3,0,-   & %m27a
-,2,41,14,36,6,0      & %0
0,1,24,14,47,14,1      & %p13a
-,0,4,7,44,39,6 & %p50a
-,0,1,2,26,52,19  %p100a
\\
& 2,74,24,0,0,-,-   & %m27b
& %0
-,0,18,13,49,19,1 & %p13b
-,0,1,3,32,50,14 & %p50b
-,-,0,0,9,46,46   %p100b
\\
N(I$\rightarrow$VIII) &
20,63,14,2,1,0,-,-  & %m27a
-,1,65,19,9,5,0,- & %0
-,1,43,29,14,13,1,- & %p13a
-,0,7,23,18,40,11,1 & %p50a
-,0,2,8,11,46,29,3   %p100a
\\
& 33,66,1,0,-,-,-,- & %m27b
&  %0
-,0,32,31,15,18,2,0   &  %p13b
-,0,3,11,14,48,23,2 & %p50b
-,-,0,1,3,30,50,15   %p100b
\\
O(I$\rightarrow$IX) &
32,52,13,2,0,-,-,-,-  & %m27a
-,1,62,28,6,2,1,-,-  & %0
-,0,36,44,11,6,2,0,- & %p13a
-,0,4,34,25,18,17,1,- & %p50a
-,-,1,12,18,22,40,8,0  %p100a
\\
& 43,56,0,-,-,-,-,-,-   & %m27b
&  %0
-,0,25,49,14,8,4,0,- &  %p13b
-,-,1,16,22,23,33,5,0 &  %p50b
-,-,0,2,6,11,52,26,2 %p100b
\\
Ne(I$\rightarrow$XI) &
5,39,52,3,0,-,-,-,-,-,-      & %m27a
-,1,60,28,10,2,0,-,-,-,-      & %0
-,0,33,38,23,6,1,0,-,-,- & %p13a
-,-,3,20,42,27,6,2,1,0,- & %p50a
-,-,0,5,24,36,19,10,6,0,-  %p100a
\\
& 8,47,45,0,-,-,-,-,-,-,- & %m27b
&  %0
-,0,22,39,29,9,1,0,-,-,-   &  %p13b
-,-,1,8,32,36,14,6,3,0,- &  %p50b
-,-,-,0,4,15,20,24,35,1,0 %p100b
\\
\noalign{\vspace{0.1cm}}
\hline

\end{tabular}
\end{center}
\footnotesize{See the caption of Table A1 for the definition of
symbols and units.}
\end{table*}
%%%%%%%%%%%%%%%%%%%%%%%%%%%%%%%%%%%%%%%%%%%%%%%%%%%%%%%%%%%%%%%%%%%%%%%%

%%%%%%%%%%%%%%%%%%%%%%%%%%%%%%% TABLE A3%%%%%%%%%%%%%%%%%%%%%%%%%%%%%%%%%
\begin{table*}
\begin{center}
\caption{Evolution of emission-line fluxes and other physical
properties for model C.} \leavevmode
\begin{tabular}[]{lccccc}
\hline
Line & April 2005 & December 2007 & March 2009 & December 2012 & December 2017\\
\hline
\noalign{\bf Model C - Spectral lines}\\ %Model 1 in lines_evol.txt
%OIV$\lambda922.7$        &
%0.016/0.014 & 0.013 & 0.013/0.012 & 0.008/0.007 & 0.005/0.003\\
%NIV$\lambda923.2$        &
%0.019/0.021 & 0.013 & 0.011/0.10 & 0.007/0      & 0.004/0.002\\
CIII$\lambda977$        &
0.349/0.308 & 0.284 & 0.247/0.223 & 0.175/0.125 & 0.116/0.069\\
NIII$\lambda991$        &
1.140/1.130 & 0.878 & 0.751/0.656 & 0.529/0.348 & 0.359/0.180\\
OVI$\lambda1032$--$1038$   &
2.360/1.180 & 2.090 & 1.920/2.270 & 1.380/2.090 & 0.880/1.620\\
%NII$\lambda1085$        &
%0.032/0.056 & 0.019 & 0.017/0.014 & 0.014/0.009 & 0.012/0.006\\
%OVI$\lambda1125$        &
%0.017/0.011 & 0.014 & 0.013/0.014 & 0.009/0.011 & 0.006/0.007\\
%NeV$\lambda1141$        &
%0.048/0.022 & 0.041 & 0.038/0.047 & 0.026/0.044 & 0.016/0.034\\
%OV$\lambda1211$         &
%0.661/0.330 & 0.648 & 0.612/0.723 & 0.457/0.655 & 0.295/0.442\\
Lyman $\alpha$ $\lambda 1216$ &
0.332/0.448 & 0.212 & 0.178/0.161 & 0.121/0.088 & 0.078/0.044\\
OV]$\lambda1218$         &
0.890/0.505 & 0.717 & 0.631/0.722 & 0.413/0.550 & 0.240/0.331\\
NV$\lambda1239,1243$    &
4.930/3.100 & 4.240 & 3.810/4.160 & 2.670/3.070 & 1.690/1.790\\
%CII$\lambda1335$    &
%0.017/0.024 & 0.014 & 0.013/0.012 & 0.012/0.009 & 0.011/0.006\\
%OV$\lambda1371$        &
%0.017/0.013 & 0.014 & 0.012/0.012 & 0.008/0.008 & 0.005/0.004\\
OIV]$\lambda1402$        &
3.850/2.590 & 3.160 & 2.730/2.740 & 1.760/1.570 & 1.030/0.736\\
%NIV$\lambda1476$        &
%0.003/0.005 & 0.002 & 0/0.001     & 0.001/0.001 & 0.001/0\\
NIV]$\lambda1486$        &
6.590/4.400 & 5.400 & 4.640/4.640 & 2.960/2.510 & 1.690/1.060\\
CIV$\lambda1549,1551$   &
1.940/1.300 & 1.570 & 1.340/1.340 & 0.842/0.694 & 0.467/0.282\\
OIII]$\lambda1661,1666$  &
1.150/1.580 & 0.733 & 0.605/0.473 & 0.439/0.230 & 0.321/0.115\\
NIII$\lambda1750$       &
3.210/4.040 & 2.200 & 1.820/1.450 & 1.280/0.659 & 0.889/0.295\\
%$[$NeIII]$\lambda1815$ &
%0.015/0.020 & 0.010 & 0.008/0.007 & 0.006/0.003 & 0.004/0.002\\
CIII]$\lambda1907,1909$ &
1.230/1.470 & 0.867 & 0.722/0.581 & 0.497/0.261 & 0.329/0.114\\
%NIII$\lambda2064$        &
%0.011/0.012 & 0.007 & 0.005/0.005 & 0.003/0.002 & 0.002/0.001\\
%NII]$\lambda2141$        &
%0.022/0.035 & 0.020 & 0.020/0.017 & 0.020/0.012 & 0.019/0.007\\
$[$OIII]$\lambda2321$       &
0.113/0.184 & 0.059 & 0.047/0.034 & 0.032/0.015 & 0.022/0.006\\
%CII]$\lambda2324-2329$   &
%0.023/0.034 & 0.020 & 0.020/0.017 & 0.019/0.012 & 0.019/0.008\\
$[$NeIV]$\lambda2424$       &
1.520/1.000 & 1.460 & 1.340/1.340 & 0.982/0.855 & 0.637/0.447\\
%$[$OII]$\lambda2471$        &
%0.019/0.039 & 0.017 & 0.017/0.013 & 0.017/0.008 & 0.016/0.004\\
%MgII$\lambda2796,2803$  &
%0.004/0.012 & 0.002 & 0.001/0.001 & 0.002/0     & 0.002/0\\
%$[$NeV]$\lambda3346$        &
%0.316/0.199 & 0.260 & 0.231/0.257 & 0.154/0.196 & 0.091/0.118\\
$[$NeV]$\lambda3426$        &
0.865/0.546 & 0.713 & 0.631/0.705 & 0.423/0.536 & 0.248/0.322\\
%$[$OII]$\lambda 3727$       &
%0.004/0.007 & 0.005 & 0.006/0.005 & 0.009/0.005 & 0.014/0.005\\
$[$NeIII]$\lambda 3869$     &
0.964/1.620 & 0.528 & 0.425/0.315 & 0.301/0.130 & 0.218/0.055\\
%FeV$\lambda3892$ &
%0.051/0.055 & 0.034 & 0.028/0.023 & 0.017/0.009 & 0.010/0.003\\
%FeII$\lambda4300$ &
%0/0         & 0     & 0/0         & 0/0         & 0/0    \\
$[$OIII]$\lambda 4363$      &
0.485/0.789 & 0.254 & 0.200/0.149 & 0.136/0.063 & 0.095/0.028\\
%H$\beta$ $\lambda 4861$&
%0.008/0.012 & 0.005 & 0.004/0.004 & 0.003/0.002 & 0.002/0.001\\
$[$OIII]$\lambda 5007$      &
9.350/16.90 & 4.630 & 3.720/2.680 & 2.730/1.150 & 2.050/0.498\\
%$[$NII]$\lambda 5755$       &
%0.013/0.024 & 0.010 & 0.010/0.008 & 0.009/0.004 & 0.007/0.002\\
%$[$OI]$\lambda 6300$       &
%0/0         & 0     & 0/0         & 0/0         & 0/0\\
H$\alpha$ $\lambda 6563$&
0.023/0.033 & 0.014 & 0.012/0.010 & 0.008/0.006 & 0.005/0.003\\
$[$NII]$\lambda 6583$       &
0.032/0.054 & 0.036 & 0.038/0.032 & 0.048/0.026 & 0.056/0.017\\
%$[$OII]$\lambda 7325$       &
%0.027/0.054 & 0.023 & 0.023/0.018 & 0.023/0.010 & 0.022/0.005\\
%SIII$\lambda9532$       &
%0.006/0.012 & 0.004 & 0/0.003     & 0.003/0.001 & 0.003/0.001\\
\noalign{\vspace{0.1cm}}
\noalign{\bf Model C - Model parameters and ionization conditions}\\
$\log{n}$ [cm$^{-3}$]
& 5.19/5.29 & 5.00 & 4.92/4.88 & 4.73/4.60 & 4.52/4.28\\
$t_{\rm nova}\,{\rm [yr]}$
& 9.2 & 11.9 & 13.2 & 16.9 & 21.9\\ 
$r_{\rm ej}\,{\rm [10^{16} cm]}$
& 4.3 & 5.6  & 6.2  & 8.0  & 10.3\\ 
$f_{\rm cov}$
& 2/3 & 2/3 & 2/3 & 2/3 & 2/3\\
$\log{U(13.6{\rm eV})}$
& -2.58/-2.68 & -2.58 & -2.57/-2.53  & -2.58/-2.45 & -2.57/-2.33\\
C(I$\rightarrow$VII) &
%0,0.7,33.5,17.1,36.7,11.0,1.0 & %m27a
-,1,34,17,37,11,1 & %m27a
%0,0.8,26.0,17.3,40.8,13.6,1.4 &%0
-,1,26,17,41,14,1 &%0
%0,1.0,24.5,16.7,41.7,14.5,1.5 & %p13a
-,1,24,17,42,15,2 & %p13a
%0,1.5,24.7,15.3,41.8,15.2,1.7 & %p50a
-,2,25,15,42,15,2 & %p50a
%0,2.3,26.1,13.5,41.0,15.3,1.8   %p100a
-,2,26,14,41,15,2   %p100a
\\
%& 0,1.2,50.2,14.1,28.3,5.8,0.4 & %m27b
& -,1,50,14,28,6,0 & %m27b
&  %0
%0,0.8,19.5,16.1,44.0,17.5,2.1 &  %p13b
-,1,20,16,44,18,2 &  %p13b
%0,1.0,13.1,11.9,44.9,25.1,4.1 &  %p50b
-,1,13,12,45,25,4 &  %p50b
%0,1.1,9.6,7.6,40.8,32.9,8.0      %p100b
-,1,10,8,41,33,8      %p100b
\\
N(I$\rightarrow$VIII) &
%0,0.2,29.7,33.9,12.0,19.8,4.0,0.3 & %m27a
-,0,30,34,12,20,4,0 & %m27a
%0,0.3,20.9,35.3,13.2,24.5,5.4,0.4 & %0
-,0,21,35,13,25,5,0 & %0
%0,0.3,19.6,34.4,13.6,25.8,5.9,0.4 & %p13a
-,0,20,34,14,26,6,0 & %p13a
%0,0.5,20.3,31.8,13.9,26.5,6.4,0.5 & %p50a
-,1,20,32,14,27,6,1 & %p50a
%0,0.9,22.9,28.9,14.1,26.1,6.7,0.5   %p100a
-,1,23,29,14,26,7,1   %p100a
\\
%& 0,0.4,48.9,27.0,10.3,11.5,1.8,0.1 & %m27b
& -,0,49,27,10,12,2,0 & %m27b
&  %0
%0,0.3,15.1,33.4,13.9,29.2,7.5,0.6 &  %p13b
-,0,15,33,14,29,8,1 &  %p13b
%0,0.3,10.2,25.7,13.8,35.6,12.9,1.3&  %p50b
-,0,10,26,14,36,13,1 & %p50b
%0,0.3,7.6,17.4,11.8,39.1,20.7,3.1    %p100b
-,0,8,17,12,39,21,3    %p100b
\\
O(I$\rightarrow$IX) &
%0,0.1,20.9,50.7,12.3,9.4,5.8,0.8,0 & %m27a
-,0,21,51,12,9,6,1,- & %m27a
%0,0.2,12.1,52.7,14.7,11.5,7.7,1.1,0 & %0
-,0,12,53,15,12,8,1,- & %0
%0,0.2,11.1,51.4,15.4,12.1,8.4,1.2,0 & %p13a
-,0,11,51,15,12,8,1,- & %p13a
%0,0.3,11.9,48.3,15.9,12.8,9.4,1.4,0 & %p50a
-,0,12,48,16,13,9,1,- & %p50a
%0,0.5,14.3,45.1,15.6,13.0,9.9,1.6,0.1 %p100a
-,1,14,45,16,13,10,2,0 %p100a
\\
%& 0,0.3,39.3,42.1,8.7,6.4,3.0,0.3,0 & %m27b
& -,0,39,42,9,6,3,0,- & %m27b
&  %0
%0,0.1,8.4,49.4,16.9,13.1,10.4,1.7,0.1 &  %p13b
-,0,8,49,17,13,10,2,0 &  %p13b
%0,0.1,5.9,39.1,19.1,15.3,16.7,3.6,0.2 &  %p50b
-,0,6,39,19,15,17,4,0 &  %p50b
%0,0.1,4.8,28.1,18.1,16.2,24.4,7.6,0.5    %p100b
-,0,5,28,18,16,24,8,1    %p100b
\\
Ne(I$\rightarrow$XI) &
%0,0.1,28.5,47.8,19.2,3.8,0.5,0.2,0 & %m27a
-,0,29,48,19,4,1,0,-,-,- & %m27a
%0,0.1,19.5,51.9,22.9,4.7,0.6,0.2,0 & %0
-,0,20,52,23,5,1,0,-,-,- & %0
%0,0.1,18.3,51.5,24.0,5.0,0.7,0.2,0.1 & %p13a
-,0,18,52,24,5,1,0,0,-,- & %p13a
%0,0.2,19.7,49.4,24.4,5.2,0.7,0.2,0.1 & %p50a
-,0,20,49,24,5,1,0,0,-,- & %p50a
%0,0.4,23.4,46.7,23.3,5.2,0.7,0.2,0.1  %p100a
0,0,23,47,23,5,1,0,0,-,-  %p100a
\\
%& 0,0.3,48.8,36.6,12.1,2.1,0.2,0.1,0 & %m27b
& -,0,49,37,12,2,0,0,-,-,- & %m27b
&  %0
%0,0.1,14.2,50.8,27.8,5.9,0.9,0.3,0.1 &  %p13b
-,0,14,51,28,6,1,0,0,-,- &  %p13b
%0,0.1,10.0,43.6,35.5,8.6,1.5,0.5,0.2 &  %p50b
0,0,10,44,35,9,2,1,0,-,- &  %p50b
%0,0.1,8.1,35.8,39.6,12.1,2.6,1.1,0.5    %p100b
-,0,8,36,40,12,3,1,1,0,-    %p100b
\\
\noalign{\vspace{0.1cm}}
\hline

\end{tabular}
\end{center}
\footnotesize{See the caption of Table A1 for the definition of
symbols and units.}
\end{table*}
%%%%%%%%%%%%%%%%%%%%%%%%%%%%%%%%%%%%%%%%%%%%%%%%%%%%%%%%%%%%%%%%%%%%%%%%

%%%%%%%%%%%%%%%%%%%%%%%%%%%%%%% TABLE A4%%%%%%%%%%%%%%%%%%%%%%%%%%%%%%%%%
\begin{table*}
\begin{center}
\caption{Evolution of emission-line fluxes and other physical
properties for model D.} \leavevmode
\begin{tabular}[]{lccccc}
\hline
Line & April 2005 & December 2007 & March 2009 & December 2012$^{(a)}$ & December 2017$^{(a)}$\\
\hline
\noalign{\bf Model D - Spectral lines}\\ %Model 1 in lines_evol.txt
%OIV$\lambda922.7$        &
%0/0         & 0.003 & 0.005/0.005 & 0/0         & 0/0\\
%NIV$\lambda923.2$        &
%0/0         & 0.002 & 0.004/0.004 & 0/0         & 0/0\\
CIII$\lambda977$        &
0/0         & 0.061 & 0.063/0.039 & 0.003/0     & 0.002/0\\
NIII$\lambda991$        &
0/0         & 0.213 & 0.208/0.138 & 0.009/0     & 0.005/0\\
OVI$\lambda1032$--$1038$   &
0/0         & 0.371 & 1.41/1.83   & 0.714/0.093 & 0.309/0\\
%NII$\lambda1085$        &
%0.005/0.002 & 0.020 & 0.005/0.002 & 0/0         & 0/0\\
%OVI$\lambda1125$         &
%0/0         & 0.002 & 0.007/0.008 & 0.001/0     & 0/0\\
%NeV$\lambda1141$         &
%0/0         & 0.007 & 0.035/0.049 & 0.028/0.004 & 0.012/0\\
%OV$\lambda1211$         &
%0/0         & 0.079 & 0.350/0.481 & 0.067/0.004 & 0.030/0\\
Lyman $\alpha$ $\lambda 1216$ &
0.391/0.411 & 0.403 & 0.191/0.145 & 0.040/0.015 & 0.023/0\\
OV]$\lambda1218$         &
0/0         & 0.149 & 0.492/0.599 & 0.050/0.003 & 0.021/0\\
NV$\lambda1239,1243$    &
0/0         & 0.746 & 2.01/2.27   & 0.155/0.013 & 0.050/0\\
%CII$\lambda1335$         &
%0.006/0.004 & 0.007 & 0.002/0.001 & 0/0         & 0/0\\
%OV$\lambda1371$         &
%0/0         & 0.002 & 0.005/0.005 & 0/0         & 0/0\\
OIV]$\lambda1402$        &
0/0         & 0.493 & 1.11/1.09   & 0.028/0.001 & 0.014/0\\
%NIV$\lambda1476$        &
%0.007/0.006 & 0     & 0.001/0     & 0/0         & 0/0\\
NIV]$\lambda1486$        &
0/0         & 0.784 & 1.72/1.65   & 0.018/0.001 & 0.009/0\\
CIV$\lambda1549,1551$   &
0/0         & 0.216 & 0.456/0.391 & 0.006/0.001 & 0.003/0\\
$[$NeIV]$\lambda1602$       &
0/0         & 0.086 & 0.161/0.142 & 0.006/0.001 & 0.003/0\\
OIII]$\lambda1661,1666$  &
0/0         & 0.296 & 0.198/0.088 & 0.003/0     & 0.002/0\\
NIII$\lambda1750$       &
0.001/0     & 0.697 & 0.496/0.251 & 0.004/0     & 0.003/0\\
%$[$NeIII]$\lambda1815$         &
%0/0         & 0.003 & 0.002/0.001 & 0/0         & 0/0\\
CIII]$\lambda1907,1909$ &
0.008/0.004 & 0.242 & 0.173/0.086 & 0.001/0     & 0.001/0\\
%NIII$\lambda2064$         &
%0/0         & 0.002 & 0.002/0.001 & 0/0         & 0/0\\
NII]$\lambda2141$        &
0.118/0.092 & 0.005 & 0.002/0.001 & 0/0         & 0/0\\
%$[$OIII]$\lambda2321$       &
%0/0         & 0.040 & 0.020/0.007 & 0/0         & 0/0\\
CII]$\lambda2324-2329$   &
0.062/0.063 & 0.005 & 0.002/0.001 & 0/0         & 0/0\\
$[$NeIV]$\lambda2424$       &
0/0         & 0.137 & 0.341/0.348 & 0.023/0.003 & 0.013/0\\
$[$OII]$\lambda2471$        &
0.283/0.200 & 0.007 & 0.002/0.001 & 0/0         & 0/0\\
MgII$\lambda2796,2803$  &
0.065/0.031 & 0.004 & 0.001/0     & 0/0         & 0/0\\
%$[$NeV]$\lambda3346$        &
%0/0         & 0.046 & 0.144/0.173 & 0.021/0.002 & 0.009/0\\
$[$NeV]$\lambda3426$        &
0/0         & 0.127 & 0.394/0.473 & 0.059/0.005 & 0.026/0\\
$[$NeI]$\lambda3467$        &
0.056/0.084 & 0     & 0/0         & 0/0         & 0/0\\
%$[$OII]$\lambda 3727$       &
%0.021/0.011 & 0.001 & 0/0         & 0/0         & 0/0\\
$[$NeIII]$\lambda 3869$     &
0.079/0.062 & 0.355 & 0.151/0.051 & 0.001/0     & 0.001/0\\
%FeV$\lambda3892$         &
%0/0         & 0.008 & 0.008/0.006 & 0/0         & 0/0\\
%FeII$\lambda4300$ &
%0.122/0.166 & 0     & 0/0         & 0/0         & 0/0\\
$[$OIII]$\lambda 4363$      &
0/0         & 0.173 & 0.088/0.031 & 0.001/0     & 0/0\\
%H$\beta$ $\lambda 4861$&
%0.007/0.009 & 0.011 & 0.005/0.004 & 0.001/0     & 0/0\\
$[$OIII]$\lambda 5007$      &
0.007/0.002 & 4.30  & 1.73/0.480  & 0.005/0     & 0.003/0\\
$[$NII]$\lambda 5755$       &
0.185/0.168 & 0.005 & 0.001/0.001 & 0/0         & 0/0\\
%FeII$\lambda 6200$       &
%0.058/0.086 & 0     & 0/0         & 0/0         & 0/0\\
$[$OI]$\lambda 6300$       &
0.348/.454 & 0      & 0/0         & 0/0         & 0/0\\
H$\alpha$ $\lambda 6563$&
0.022/0.030 & 0.031 & 0.014/0.010 & 0.002/0.001 & 0.001/0\\
$[$NII]$\lambda 6583$       &
0.244/0.159 & 0.013 & 0.003/0.001 & 0/0     & 0/0\\
$[$OII]$\lambda 7325$       &
0.375/0.264 & 0.010 & 0.003/0.001 & 0/0     & 0/0\\
%SIII$\lambda9532$       &
%0.003/0.002 & 0.006 & 0.001/0     & 0/0     & 0/0\\
\noalign{\vspace{0.1cm}}
\noalign{\bf Model D - Model parameters and ionization conditions}\\
$\log{n}$ [cm$^{-3}$]
& 6.20/6.55 & 5.50 & 5.30/5.20 & 4.46/4.15 & 4.11/3.63\\
$t_{\rm nova}\,{\rm [yr]}$
& 1.5 & 4.2 & 5.5 & 9.2 & 14.2\\ 
$r_{\rm ej}\,{\rm [10^{16} cm]}$
& 0.7 & 2.0 & 2.6 & 4.3$^{(a)}$ & 6.7$^{(a)}$\\ 
$f_{\rm cov}$
& 0.02 & 0.10 & 0.160 & 2/3$^{(a)}$ & 2/3$^{(a)}$\\
$\log{U(13.6{\rm eV})}$
& -3.28/-3.63 & -2.58 & -2.38/-2.28  & -1.76/-1.44 & -1.76/-1.28\\
C(I$\rightarrow$VII) &
41,57,2,-,-,-,- & %m27a
-,2,65,8,20,5,0 & %0
-,0,20,12,40,22,4 & %p13a
-,-,0,0,4,30,66 & %p50a
-,-,0,0,5,31,63   %p100a
\\
& 59,40,1,-,-,-,- & %m27b
& %0
-,0,7,10,42,33,8 &  %p13b
-,-,-,-,1,15,85&  %p50b
-,-,-,-,-,3,97   %p100b
\\
N(I$\rightarrow$VIII) &
74,26,0,-,-,-,-,- & %m27a
-,1,68,15,7,8,2,0 & %0
-,0,21,27,12,28,10,1 & %p13a
-,-,0,0,0,9,42,48 & %p50a
-,-,0,0,0,9,41,48  %p100a
\\
& 80,20,-,-,-,-,-,- & %m27b
&  %0
-,-,7,25,12,36,17,2 &  %p13b
-,-,-,-,-,2,23,75 &  %p50b
-,-,-,-,-,0,5,95   %p100b
\\
O(I$\rightarrow$IX) &
70,30,0,-,-,-,-,-,- & %m27a
-,1,64,23,6,4,3,0,- & %0
-,0,18,40,16,11,12,3,0 & %p13a
-,-,0,1,1,2,21,49,26 & %p50a
-,-,0,1,1,2,20,49,26  %p100a
\\
& 76,24,-,-,-,-,-,-,- & %m27b
&  %0
-,-,4,39,19,13,18,6,0 &  %p13b
-,-,-,-,0,0,6,38,56 &  %p50b
-,-,-,-,-,-,0,10,90   %p100b
\\
Ne(I$\rightarrow$XI) &
61,30,10,-,-,-,-,-,-,-,- & %m27a
-,1,67,21,10,2,0,0,-,-,- & %0
-,0,20,40,31,8,1,1,0,-,- & %p13a
-,-,0,3,15,21,14,11,27,9,1 & %p50a
-,-,1,3,15,20,13,11,28,9,1  %p100a
\\
& 71,23,6,-,-,-,-,-,-,-,- & %m27b
&  %0
-,-,6,38,40,12,3,1,1,0,- &  %p13b
-,-,0,0,2,5,5,6,39,35,6 &  %p50b
-,-,-,-,-,-,-,0,8,46,46   %p100b
\\
\noalign{\vspace{0.1cm}}
\hline

\end{tabular}
\end{center}

\footnotesize{See the caption of Table A1 for the definition of
symbols and units.\\
{\it Note}: $^{(a)}$: In this model, the 2012 and 2017 epochs have $r_{\rm
ej}>d$ (the distance between the XRS and the nova), i.e. the XRS
becomes engulfed in the expanding nova shell. Then, for these epochs
we assume $f_{\rm cov}=2/3$, and use $r_{\rm ej}$ as the distance from
the XRS to the shell; this is quite inaccurate, but the emission lines
remain very weak also for more detailed calculations.}
\end{table*}
%%%%%%%%%%%%%%%%%%%%%%%%%%%%%%%%%%%%%%%%%%%%%%%%%%%%%%%%%%%%%%%%%%%%%%%%

\end{document}